\documentclass[10pt,aps,prb,superscriptaddress,nobibnotes,notitlepage,twocolumn]{revtex4-2}

\usepackage{mathptmx}
\usepackage{mathtools}
\usepackage[normalem]{ulem}
\usepackage{graphicx}
\usepackage{multirow}
\usepackage{xcolor}

\usepackage{hyperref}
\hypersetup{colorlinks=true, linkcolor=blue, citecolor=blue, urlcolor=blue}

\newcommand{\beq}[1]{\begin{equation}\label{#1}}
\newcommand{\eeq}{\end{equation}}
\newcommand{\eep}{\;.\end{equation}}
\newcommand{\eec}{\;,\end{equation}}

\newcommand{\ep}{\epsilon}

\DeclareMathAlphabet{\mathcal}{OMS}{cmsy}{m}{n}

\newcommand{\N}{\mathcal{N}}

\newcommand{\bvec}[1]{\mathbf{#1}}

\newcommand{\kv}{\bvec{k}}

\newcommand{\bv}{\bvec{b}}
\newcommand{\rv}{\bvec{r}}
\newcommand{\R}{\bvec{R}}

\newlength{\characterscolwidth}
\newcommand{\characterscell}[1]{\makebox[\characterscolwidth][c]{#1}}
\newlength{\charlabelwidth}
\newcommand{\charlabelcell}[1]{\makebox[\charlabelwidth][c]{#1}}
\newlength{\nnlabelcolwidth}
\newlength{\nnvaluecolwidth}
\newcommand{\nnlabelcell}[1]{\makebox[\nnlabelcolwidth][c]{#1}}
\newcommand{\nnvaluecell}[1]{\makebox[\nnvaluecolwidth][c]{#1}}

\newcommand*\dd{\mathop{}\!\mathrm{d}}

\newcommand{\HarvardQSE}{Quantum Science and Engineering, Harvard University, Cambridge, Massachusetts 02138, USA}
\newcommand{\HarvardPhysics}{Department of Physics, Harvard University, Cambridge, Massachusetts 02138, USA}
\newcommand{\HarvardSeas}{John A.~Paulson School of Engineering and Applied Sciences, Harvard University, Cambridge, Massachusetts 02138, USA}
\newcommand{\FSU}{Department of Physics, Florida State University, Tallahassee, Florida 32306, USA}
\newcommand{\NTU}{School of Electrical and Electronic Engineering, Nanyang Technological University Singapore, 50 Nanyang Avenue, 639798, Singapore}

\begin{document}

\title{
On the nature of low-energy electronic states in twisted bilayer graphene: 
a first-principles study
}

\author{Albert Zhu}
\email{axzhu@g.harvard.edu}
\affiliation{\HarvardQSE}

\author{Daniel Bennett}
\affiliation{\HarvardSeas}
\affiliation{\NTU}

\author{Daniel T.~Larson}
\affiliation{\HarvardPhysics}

\author{Mohammed M. Al Ezzi}
\affiliation{\HarvardSeas}

\author{Efstratios Manousakis}
\affiliation{\FSU}

\author{Efthimios Kaxiras}
\affiliation{\HarvardSeas}
\affiliation{\HarvardPhysics}

\begin{abstract}
The nature of low-energy electronic states in twisted bilayer graphene (TBG) near the magic angle has been extensively discussed in the literature on qualitative grounds, since these states play a key role in the construction of effective few-band models toward a many-body theory.  Here, we report extensive and systematic first-principles calculations of TBG electronic states in the magic-angle range, 
using density functional theory with no further approximations. 
We construct moir\'e-scale localized Wannier states for the low-energy bands and develop an effective ten-band model for each $K$ valley. 
The Wannier orbitals contain both atomic and moir\'e-scale details and are centered near the high-symmetry stackings in the moir\'e cell.
We calculate Coulomb and exchange integrals of these localized states, quantities that are the basic ingredients in many-body models. We also compare the values of these integrals to those obtained or assumed in previous works.
By symmetrizing the ten-band Hamiltonians and basis orbitals, we provide insight into how the dominant hoppings between the flat-band orbitals evolve as a function of twist angle.
\end{abstract}

\maketitle


\section{Introduction}
Since the observation of numerous exotic phenomena in twisted bilayer graphene (TBG) near the magic angle, 
including superconductivity~\cite{cao2018unconventional,lu2019superconductors,yankowitz2019tuning,oh2021evidence,tanaka2025superfluid,banerjee2025superfluid,codecido2019correlated}, 
correlated insulating behavior~\cite{cao2018correlated,tian2024dominant}, 
and topological phases~\cite{sharpe2019emergent,serlin2020intrinsic,nuckolls2020strongly,wu2021chern,choi2021correlation,xie2021fractional,stepanov2021competing},
recent experimental advances have begun to shed light on the origin of these phenomena with increasing resolution.
Scanning tunneling microscopy (STM) can probe the spatial variation of wavefunctions in TBG with atomic-scale resolution, 
revealing symmetry-broken textures thought to be related to correlated states and superconductivity~\cite{nuckolls2023quantum}.
Quantum twisting microscopy can directly measure the electronic band structure of twisted interfaces~\cite{inbar2023quantum} and has
revealed that the flat bands of TBG under strong electron correlations are drastically changed from their non-interacting counterparts~\cite{xiao2026imaging}.
Related optical probes have also begun to access the role of quantum geometry and interactions in the flat bands~\cite{krishna2025terahertz}.

Understanding these experimental observations requires a rigorous, microscopic, single-particle model describing the bands of interest that can serve as the foundation for many-body models.
A traditional approach to obtaining such a fundamental model has been to derive reduced tight-binding models of the low-energy bands from single-particle Hamiltonians of TBG, such as 
continuum models~\cite{dos2007graphene,Bistritzer2011,lopes2012continuum,carr2019exact,fang2019angle,guinea2019continuum,kang2023pseudomagnetic,vafek2023continuum,miao2023truncated,ezzi2024analytical}
or parameterized tight-binding models~\cite{trambly2010localization,jung2014accurate,fang2016electronic,carr2017twistronics,bernevig2021twisted,bennett2024twisted}.
The resulting basis orbitals of such reduced models can then be used to calculate electronic interactions~\cite{koshino2018maximally,kang2018symmetry,carr2019derivation,calderon2020interactions,song2022magic}.
However, early works isolating only the flat bands of TBG encountered the Wannier obstruction associated with the fragile topology of TBG~\cite{zou2018band,po2018origin,po2019faithful,ahn2019failure,song2019all},
resulting in highly delocalized Wannier orbitals that are not ideal for modeling local interactions.
Moreover, models derived from continuum Hamiltonians or parameterized tight-binding descriptions do not contain the full atomic-scale wavefunctions or the effect of atomic relaxation as obtained from realistic first-principles treatment.

In this work, we build upon the DFT framework of Refs.~\cite{bennett2025accurate,zhu2026wavefunction,zhu2026twisted} to derive reduced 
tight-binding models and real-space Wannier orbitals for magic-angle TBG entirely from first-principles calculations.
Starting from fully relaxed DFT structures, we Wannierize the DFT bands within a manifold closest to the charge neutrality point (CNP), which is separated from the rest of the spectrum by a $\kv$-dependent gap.
The resulting tight-binding Hamiltonian naturally splits into two blocks, which we identify with the two valleys of TBG. 
This approach provides a valley-resolved, \emph{ab initio} model that retains the spatial information contained in the DFT wavefunctions while reducing the electronic structure to a lower-dimensional manifold for further analysis.
The corresponding Wannier orbitals exhibit both atomic and moir\'e-scale character, with centers and weight distributed near the AA, domain-wall (DW), and AB/BA regions of the moir\'e unit cell.

Using these orbitals as a basis, we compute on-site and nearest-neighbor Coulomb interactions, as well as exchange interactions, as a function of twist angle.
To contextualize our findings and compare with previous work, we project the reduced models into a symmetrized basis commonly used in related tight-binding models of TBG.
This process reveals a dramatic twist-angle dependence in certain orbital hoppings relevant to the dispersion of the flat bands.
The results provide a route from \emph{ab initio} electronic structure to reduced, rigorous tight-binding models of magic-angle TBG 
and offer the necessary ingredients for modeling of electron correlations in the context of a many-body theory completely derived from first principles.

\begin{figure}[t]
\centering
\includegraphics[width=1.0\columnwidth]{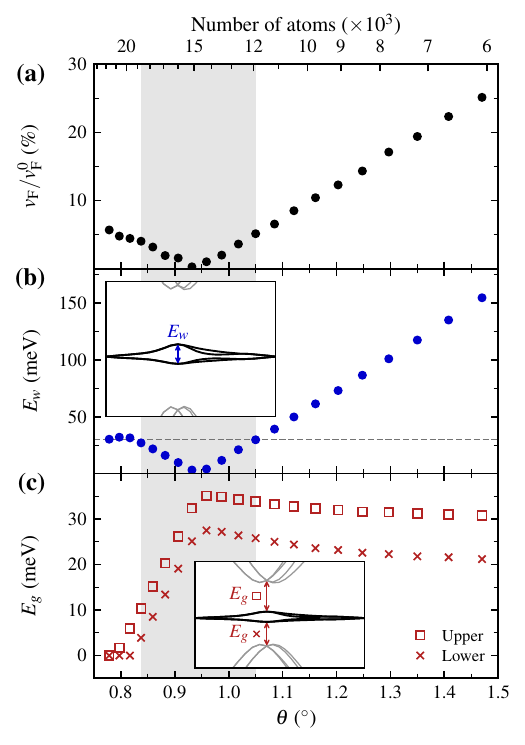}
\caption{%
Features of the DFT electronic bands in TBG as a function of twist angle $\theta$.
{\bf (a)} Fermi velocity $v_{\rm F}$ of the Dirac cones at the $K$ point in the moir\'e Brillouin zone, normalized by the Fermi velocity of monolayer graphene $v_{\rm F}^0$.
{\bf (b)} Bandwidth $E_w$, defined as the energy difference between the highest and lowest energy states in the flat bands, as shown in the inset.
The dashed line at 30 meV denotes the bandwidth cutoff we use to define the magic range in the DFT calculations.
{\bf (c)} Band gaps $E_g$ between the flat bands and the remote bands. The upper (lower) band gap is defined as the energy difference between the upper 
(lower) flat band and its neighboring remote band at the $\Gamma$ point, as shown in the inset.
The shaded region in each panel denotes the magic range.
The bottom axis gives the twist angle $\theta$ in degrees, while the top axis shows the corresponding number of atoms per moir\'e cell.
}
\label{fig:bands}
\end{figure}


\begin{figure*}[t]
\centering
\includegraphics[width=1.0\textwidth]{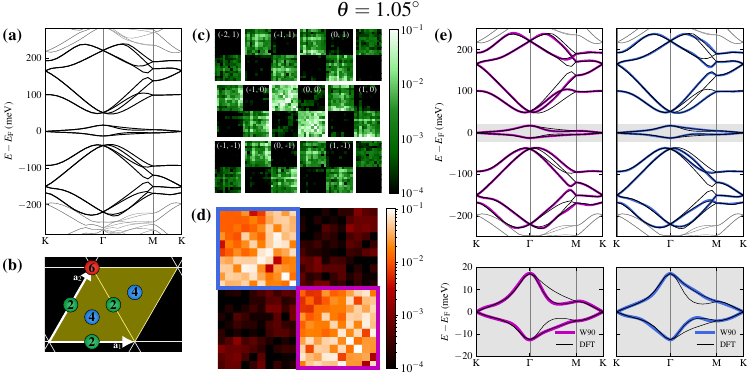}
\caption{%
Wannierization procedure and resulting tight-binding model for $\theta = 1.05^\circ$.
{\bf (a)} Band structure of TBG in the magic range obtained from DFT, with the $20$ bands selected for Wannierization in black. 
{\bf (b)} The $20$ initial orbital centers used for Wannierization. Intersections of white lines denote the AA regions.
$6$ orbitals (red) are centered on a triangular lattice (AA region), $6$ orbitals (green) are centered on a Kagome lattice (DW) with $2$ per site, and $8$ orbitals (blue) are centered on a honeycomb lattice (AB/BA) with $4$ per site.
{\bf (c)} Amplitude of the real-space hopping matrix elements $\vert H_{ij}(\mathbf{R}) \vert$, in units of eV.
Each element $H_{ij}(\mathbf{R})$ denotes hopping from orbital $j$ in the unit cell shifted by $\mathbf{R} = n\mathbf{a}_1 + m\mathbf{a}_2 \equiv (n,m)$ to orbital $i$ in the original cell.
{\bf (d)} Momentum-dependent tight-binding Hamiltonian $H(\mathbf{k})$ assembled from $H_{ij}(\mathbf{R})$ via Eq.~\eqref{eq:HRtoHk} for $\mathbf{k} = \mathrm{K}$.
{\bf (e)} Band structures obtained for individual valleys by diagonalizing the $10 \times 10$ blocks of $H(\mathbf{k})$ outlined in blue and magenta in {\bf (d)}.
}
\label{fig:wannierization}
\end{figure*}

\section{Methods
\label{sec:methods}
}

Fully relaxed atomic structures, electronic bands, and Bloch states of TBG were calculated using the \textsc{Siesta} DFT code~\cite{soler2002siesta} for commensurate twist angles between $21.79^\circ$ and $0.78^\circ$, with a summary of the electronic bands shown in Fig.~\ref{fig:bands} (for details of the DFT calculations, see Refs.~\cite{zhu2026wavefunction,zhu2026twisted}).
For commensurate angles $\theta$ in the range $[0.93^\circ, 1.41^\circ]$, Wannier orbitals were obtained using the \textsc{Wannier90} package~\cite{marzari1997maximally,mostofi2008wannier90} by minimizing the total spread $\Omega$ of $N$ Wannier orbitals $|w_n(\rv)\rangle$:
\begin{equation}
    \Omega = \sum_{n=1}^N \Omega_n = \sum_{n=1}^N \left[ \langle r^2 \rangle_n - \vert \langle \rv \rangle_n \vert^2 \right],
\label{eq:spread}
\end{equation}
where $\langle r^2 \rangle_n = \langle w_n(\rv) \vert r^2 \vert w_n(\rv) \rangle$, and $\langle \rv \rangle_n = \langle w_n(\rv) \vert \rv \vert w_n(\rv) \rangle$.
We select the bands to be included in the summation by observing that there exists a manifold of $N=20$ states (10 above and 10 below the CNP), which are separated from the rest of the energy spectrum by a $\kv$-dependent band gap, as shown in Fig.~\ref{fig:wannierization} (a).
As such, disentanglement of these bands from the rest is not required.
The spread $\Omega$ can be written in terms of the overlaps $M_{mn}^{\kv, \bv} = \langle u_{m,\kv} \vert u_{n,\kv + \bv} \rangle$ between periodic parts of the Bloch states $u_{m,\kv}(\rv)$ on a discrete $\kv$-point grid~\cite{marzari1997maximally}.
We calculate the Bloch states $\psi_{m,\kv}(\rv) = e^{i\kv \cdot \rv}u_{m,\kv}(\rv)$ on a uniform $6 \times 6$ $\kv$-point grid, the minimal grid that includes the high-symmetry points $\Gamma$, K, and M in the moir\'e Brillouin zone (mBZ).
The overlaps $M_{mn}^{\kv, \bv}$ are then used to determine the unitary matrices $U^{(\kv)}$ for constructing each $w_{n,\bvec{R}}(\rv)$ from the Bloch states:
\begin{equation}
    w_{n,\bvec{R}}(\rv) = \frac{1}{N_k} \sum_{\kv} \left[\sum_m U_{mn}^{(\kv)} \psi_{m,\kv}(\rv) \right] e^{-i\kv \cdot \bvec{R}},
\end{equation}
where $N_k$ is the number of $\kv$-points in the summation.
We use a set of trial orbitals $g_n(\rv)$ as an initial guess for the Wannier orbitals.
Motivated by the choice of orbitals in previous ten-band models~\cite{po2019faithful,carr2019exact}, we choose $20$ such $g_n(\rv)$ with atomic $p_z$ character with centers in the moir\'e cell shown in Fig.~\ref{fig:wannierization} (b): 6 at the AA regions (corresponding to a triangular lattice); 2 at each of the three DW centers (forming a Kagome lattice); and 4 in each of the AB and BA regions (forming a honeycomb lattice).
The Wannierization procedure is then iterated until $\Omega$ converges.

\begin{figure*}[t]
\centering
\includegraphics[width=\textwidth]{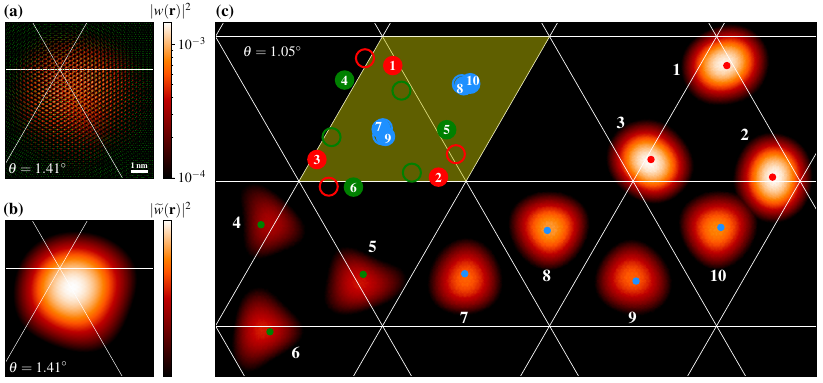}
\caption{%
Wannier orbitals from first principles.
{\bf (a)} Probability density $\vert w(\rv) \vert^2$ integrated along the out-of-plane direction for a single Wannier orbital $w(\rv)$ for $\theta = 1.41^\circ$.
Green dots indicate atomic positions.
{\bf (b)} Convolution of $\vert w(\rv) \vert^2$ in (a) with a Gaussian with a standard deviation of $10$ \AA.
{\bf (c)} Shapes and locations of all 10 Wannier orbitals from one valley for $\theta = 1.05^\circ$.
Yellow parallelogram in the top left denotes the moir\'e cell containing circles that denote the centers of the orbitals within the cell.
Closed circles denote orbital centers in one valley; open circles denote orbital centers in the other valley.
The colors of the circles correspond to the classification of each orbital as described in the text.
The probability densities of orbitals 1 through 10 are shown.
}
\label{fig:orbitals}
\end{figure*}


\section{Results}
\subsection{Electronic band structure features
\label{sec:bands}
}

We calculate the electronic bands of TBG for commensurate twist angles as small as $0.78^\circ$, corresponding to over 20,000 atoms per moir\'e cell, from first-principles DFT.
In Fig.~\ref{fig:bands} we show the Fermi velocity $v_{\rm F}$, bandwidth $E_w$ of the flat bands, and band gaps $E_g$ between the flat bands and the remote bands as a function of twist angle $\theta$.
The Fermi velocity and bandwidth both attain a minimum between $0.93^\circ$ and $0.96^\circ$ ($v_{\rm F}/v_{\rm F}^0 = 0.3$\% and $E_w = 3$ meV at $0.93^\circ$), while the band gaps reach a maximum in that range (upper $E_g = 35$ meV and lower $E_g = 28$ meV at $0.96^\circ$).
At an angle $\theta^*$ between $0.93^\circ$ and $0.96^\circ$, the flat bands touch at the $\Gamma$ point, exchange wavefunction character, and then cross along the $\Gamma$--M path as the twist angle decreases~\cite{bennett2024twisted,zhu2026wavefunction}. 
(We note that the flat bands do not cross each other at the Dirac points.)
We thus define the magic angle in the \textsc{Siesta} DFT calculations to be $0.96^\circ$, the commensurate angle above $\theta^*$ with the smallest Fermi velocity and bandwidth.
We note that the precise value of the magic angle calculated from DFT depends on the choice of exchange-correlation functional and van der Waals interactions~\cite{zhu2026twisted};
however, the Fermi velocity, bandwidth, and band gaps calculated from DFT are consistent with other models aside from a small offset in twist angle~\cite{zhu2026twisted}.

In our case, the magic angle in the \textsc{Siesta} DFT calculations is somewhat smaller than the value of the magic angle typically used in prior literature ($\sim 1.05^\circ$), 
the angle at which the Fermi velocity vanishes in the Bistritzer--MacDonald model~\cite{Bistritzer2011}.
However, because the Fermi velocity never truly vanishes in models accounting for atomic relaxation, such as our DFT calculations and Refs.~\cite{carr2019exact,bennett2024twisted}, we also define a ``magic range''
where the flat bands remain isolated and possess a small enough bandwidth ($\sim 30$ meV or less) such that electronic interactions become important.
In Fig.~\ref{fig:bands} (c), the smallest angle at which the flat bands are isolated is $0.84^\circ$; at that angle, the bandwidth of the flat bands is $\sim 30$ meV.
The largest angle at which the bandwidth remains below $\sim 30$ meV is $1.05^\circ$. 
We thus define $[0.84^\circ, 1.05^\circ]$ as the magic range of our calculations, as shown by the shaded region in Fig.~\ref{fig:bands}.
The width of our magic range is in agreement with the range of angles, $0.93^\circ$ to $1.16^\circ$, where experimental studies report correlated phases and superconductivity~\cite{cao2018unconventional,codecido2019correlated}.

\subsection{Tight-binding models}

The tight-binding Hamiltonian in the basis of the localized orbitals obtained during Wannierization is specified by the matrix elements for hopping 
from orbital $j$ in the unit cell located at $\mathbf{R} = n\mathbf{a}_1 + m\mathbf{a}_2 \equiv (n,m)$ to orbital $i$ in the original cell, $\mathbf{R} = (0,0)$, where $\mathbf{a}_1$ and $\mathbf{a}_2$ are the moir\'e lattice vectors.
A representative tight-binding model is shown in Fig.~\ref{fig:wannierization} (c) and (d) for $\theta = 1.05^\circ$.
In Fig.~\ref{fig:wannierization} (c) we show the $20\times20$ blocks $H_{ij}^{(n,m)}$ located at the $(n,m)$ moir\'e cell corresponding to the initial orbital $j$.
The squares at position $i,j$ within each $20\times20$ block are colored based on the magnitude of the hopping from $w_{j,(n,m)}$ to $w_{i,(0,0)}$.
The $20 \times 20$ matrix of hoppings can be block-diagonalized into two $10 \times 10$ blocks, with all elements in the off-diagonal blocks less than 1 meV, that is, less than $1$\% of the largest matrix element, in magnitude.
The momentum-dependent tight-binding Hamiltonian $H(\kv)$ is constructed as follows:
\begin{equation}
    H_{ij}(\kv) = \sum_{\R}H_{ij}(\R)\frac{1}{g_{\R}}\sum_{\bvec{T}}e^{i\kv \cdot(\R + \bvec{T})},
    \label{eq:HRtoHk}
\end{equation}
where $\bvec{T}$ is a correction vector for each $\R$ used by \textsc{Wannier90} to account for finite hopping distances and $g_\R$ is the number of $\bvec{T}$ vectors for each $\R$ shell.
After this transformation, the off-diagonal blocks of $H(\kv)$ remain small in comparison to the on-diagonal blocks.
We thus neglect the off-diagonal blocks of $H(\kv)$ and work with the two $10\times 10$ on-diagonal blocks.

We compute the eigenvalues of the two $10 \times 10$ on-diagonal blocks of $H(\kv)$ along the K--$\Gamma$--M--K path in the mBZ, yielding the band structures shown in Fig.~\ref{fig:wannierization} (e).
We observe that diagonalizing the $10 \times 10$ block outlined in blue in Fig.~\ref{fig:wannierization} (d) results in the band structure highlighted in blue in Fig.~\ref{fig:wannierization} (e) that closely reproduces 10 of the 20 DFT bands, corresponding to one $K$ valley.
Similarly, diagonalizing the block outlined in magenta closely reproduces the remaining 10 bands, corresponding to the other $K$ valley, as seen in Fig.~\ref{fig:wannierization} (e).
By comparing these results with continuum models~\cite{Bistritzer2011,carr2019exact}, which are parameterized for a single valley, we verify that each $10 \times 10$ on-diagonal block corresponds to one $K$ valley.
Consequently, each $10 \times 10$ block can be treated as a ten-band model of TBG, and all subsequent analysis will be done for a single valley using a single ten-band model.
The agreement between the bands from the effective model Hamiltonian and the DFT results is of similar quality for all other angles, especially for the bands closest to the CNP. 
When the Wannierization process results in poorer agreement, only the bands farthest away from the CNP are affected, which does not have any impact on the usefulness of the model.


\begin{figure}[t]
\centering
\includegraphics[width=1.0\columnwidth]{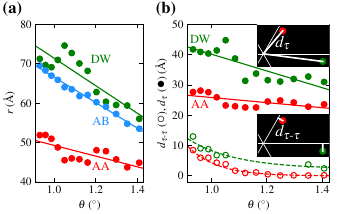}
\caption{%
Properties of the Wannier orbitals.
{\bf (a)} Radius $r_n = \sqrt{\Omega_n}$ of the orbitals as a function of $\theta$.
{\bf (b)} Distance of the AA and DW Wannier centers from the AA stacking site ($d_\tau$, filled circles) and from the nearest domain wall ($d_{\tau\text{-}\tau}$, open circles). Both distances are illustrated in the insets and plotted as a function of $\theta$.
}
\label{fig:orbital-analysis}
\end{figure}

\subsection{Properties of Wannier orbitals}

The basis states for our ten-band models are the twist-angle-dependent real-space Wannier orbitals; some examples are shown in Fig.~\ref{fig:orbitals}. In 
Fig.~\ref{fig:orbitals} (a) we show the atomic-scale probability density $\vert w(\rv) \vert^2$ of a representative orbital $w(\rv)$  for $\theta = 1.41^\circ$.
Close to the center of the Wannier orbital (near the AA region), $\vert w(\rv) \vert^2$ is a local maximum at the positions of the atoms.
In contrast, near the edge of the Wannier orbital (closer to the AB/BA and DW regions), $\vert w(\rv) \vert^2$ reaches a local maximum along bonds in between atoms.
To obtain a moir\'e-scale description of the orbital, we perform a convolution of $\vert w(\rv) \vert^2$ with a Gaussian 
$\N_{\rv}(\rv')$ centered at $\rv$:
\beq{}
\vert\widetilde{w}_{n\R}(\rv)\vert^2 \equiv (\vert w_{n\R} \vert^2 * \N)(\rv) = \sum_{\rv'} \vert w_{n\R}(\rv') \vert^2 \N_{\rv}(\rv')
\eep
The corresponding moir\'e-scale orbital is shown in Fig.~\ref{fig:orbitals} (b), while in Fig.~\ref{fig:orbitals} (c) we show the shape and location of all 10 moir\'e-scale orbitals for $\theta = 1.05^\circ$.

We classify the Wannier orbitals into three categories,
based on the spread and center position: (i) AA orbitals (labeled 1--3 in Fig.~\ref{fig:orbitals} (c)), (ii) DW orbitals (labeled 4--6), and (iii) AB/BA orbitals (2 in each region, labeled 7, 9 and 8, 10, respectively).
The dot within each orbital denotes its center and is color-coded red (AA), green (DW), or blue (AB/BA). 
The top left corner of Fig.~\ref{fig:orbitals} (c) shows
how the orbitals for each valley are centered within a single moir\'e unit cell, with filled and open circles denoting the centers of the orbitals corresponding to the two valleys.

For further analysis of the orbital character, we show in Fig.~\ref{fig:orbital-analysis} (a) the radius $r_n$ of each orbital type, defined as $r_n = \sqrt{\Omega_n}$ with $\Omega_n$ defined in Eq.~\eqref{eq:spread}.
The AA orbitals have the smallest radius, $r_n \sim 45$ \AA{} for $\theta=1.41^\circ$, and it increases only slightly as the twist angle decreases.
The AB/BA and DW orbitals have larger radii, $r_n \sim 55$ \AA{} for $\theta=1.41^\circ$, that increase more rapidly to 
$r_n\gtrsim 70$ \AA{}, as the twist angle decreases from $1.41^\circ$ to $0.93^\circ$.  In Fig.~\ref{fig:orbital-analysis} (b) 
we plot the distances of the AA and DW orbital centers
from the high-symmetry stacking regions of the moir\'e cell.
The distance $d_\tau$ of the AA and DW orbitals from the triangular lattice sites (exact AA-stacking regions) generally increases as the twist angle decreases, with the trend being more pronounced for the DW orbitals.
The offset $d_{\tau\text{-}\tau}$ of the AA and DW orbitals from the nearest domain wall, although smaller than $d_{\tau}$, shows similar trends.  

\begin{figure}[h!]
\centering
\includegraphics[width=0.9\columnwidth]{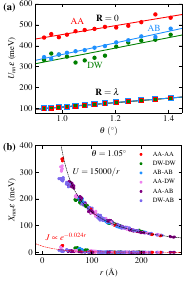}
\caption{%
Coulomb and exchange interactions in the ten-band model.
{\bf (a)} On-site (circles) and nearest-neighbor (squares) Coulomb interactions as a function of twist angle. 
Nearest-neighbor interactions are calculated between the same Wannier orbital in neighboring moir\'e cells.
{\bf (b)} Coulomb interactions $U_{nm}\epsilon$ and exchange interactions $J_{nm}\epsilon$ (denoted collectively by $X_{nm}\epsilon$, with $X = U$ or $J$) between all pairs of distinct orbitals as a function of the distance $r$ between the orbitals, all for $\theta=1.05^\circ$.
Colors correspond to the orbital types involved in the interaction.
The black dashed curve fit to the Coulomb interactions is proportional to $1/r$, 
while the red dashed curve fit to the exchange interactions is proportional to $e^{-0.024r}$ (with $r$ in units of \AA).
}
\label{fig:U}
\end{figure}

\subsection{On-site and nearest-neighbor interactions
\label{sec:U}
}

We use the Wannier orbitals to calculate Coulomb and exchange interactions in our ten-band model. 
The Coulomb interaction parameters $U$ and exchange interaction parameters $J$ are given by
\beq{eq:Hubbard}
U_{nm}(\mathbf{R}) = \frac{e^2}{4\pi\ep_{0}}\iint
\frac{\vert w_{n,\bf{0}}(\rv)\vert^2\vert w_{m,\bf{R}}(\rv')\vert^2}{\ep\vert \rv - \rv'\vert}\dd\rv\dd\rv'
\eec
and
\beq{eq:Exchange}
J_{nm}({\bf R}) = \frac{e^2}{4\pi\ep_{0}}\iint
\frac{w_{n,\bf{0}}^*(\rv) w_{m,\bf{R}}^*(\rv') w_{m,\bf{R}}(\rv) w_{n,\bf{0}}(\rv')}{\ep\vert \rv - \rv'\vert}\dd\rv\dd\rv'
\eec
respectively, where $w_{n,\bf{R}}(\rv)$ is a Wannier orbital in a moir\'e cell located at a lattice vector $\bvec{R}$ and $\epsilon$ is the dimensionless dielectric function that captures the Coulomb screening by electrons in remote bands.
In this work, we consider on-site and nearest-neighbor interactions; thus, we consider $|\bvec{R}| = 0$ or $|\bvec{R}| = \lambda$, where $\lambda$ is the moir\'e length.
Furthermore, we remain agnostic regarding the value of the dielectric function $\epsilon$ since a range of values from $\epsilon \sim 4$ to $12$ have been used in the literature~\cite{kang2019strong,calderon2020interactions,goodwin2019attractive,goodwin2019twist,koshino2018maximally}.
In general, the value of $\epsilon$ varies spatially and depends on specific experimental conditions, such as encapsulation with hexagonal boron nitride.
Here we treat $\epsilon$ as a constant and report values of $U\epsilon$ and $J\epsilon$.
A careful first-principles calculation of $\epsilon$ is an important next step.

Fig.~\ref{fig:U} (a) shows the on-site
and nearest-neighbor Coulomb interactions, $U_{nn}(0)$ and $U_{nn}(\lambda)$, for each orbital type.
The AA orbitals exhibit the largest Coulomb interactions, with $U_{nn}(0)\epsilon$ in the range of 450 to 550 meV for twist angles between $0.93^\circ$ and $1.41^\circ$.
The on-site interactions in the AB/BA and DW orbitals are somewhat weaker.
All on-site $U_{nn}(0)$ values decrease roughly linearly as the twist angle decreases, with the interactions of the AA orbitals decreasing more slowly than the other orbital types.
The nearest-neighbor interactions in Fig.~\ref{fig:U} (a) are computed for the same Wannier orbital located in nearest-neighbor moir\'e cells.
They are generally $1/4$ to $1/3$ the size of the on-site interactions and almost identical for the different orbital types, indicating that all Wannier orbitals have decayed significantly and to a similar extent after one moir\'e length.

Fig.~\ref{fig:U} (b) shows the Coulomb interactions between orbitals of different types as a function of the distance $r$ between the orbital centers for a single twist angle $\theta = 1.05^\circ$.
The Coulomb interactions decrease proportionally with $1/r$ when $r$ is greater than half of the moir\'e length $\lambda$, \emph{independent} of the twist angle.
In fact, for all twist angles between $0.93^\circ$ and $1.41^\circ$, the Coulomb interactions decrease with the same $\sim 15000/r$ scaling (with $r$ in units of \AA),
indicating that the magnitude of interactions between different orbitals produced by \textsc{Wannier90} is determined predominantly by the distance separating them, 
not the twist angle or orbital type.

The exchange interactions between orbitals of different types are also shown in Fig.~\ref{fig:U} (b).
All exchange interactions are on the order of $J\epsilon \sim 10$ meV or smaller, less than $5$\% of the on-site Coulomb interactions.
Furthermore, they decrease exponentially as the distance between the orbitals increases.

\begin{figure}[t]
\centering
\includegraphics[width=\columnwidth]{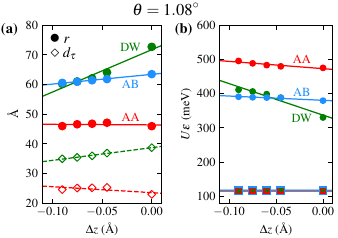}
\caption{%
Wannierization of compressed TBG layers for a twist angle of $1.08^\circ$.
{\bf (a)} Radius (solid lines) and distance from AA site (dashed lines) of different Wannier orbitals as a function of compression.
{\bf (b)} On-site (circles) and nearest-neighbor (squares) Coulomb interactions of different orbitals as a function of compression.
}
\label{fig:characterExchange}
\end{figure}

\subsection{Electronic states with increased interlayer interactions}
Finally, we study the effect of increased interlayer interactions on the Wannier states, achieved via compression of the layers of TBG using the same procedure as Ref.~\cite{zhu2026wavefunction}.
Letting $z$ denote the mean distance between the two layers, we define $\Delta z$ as the amount by which $z$ is perturbed from its equilibrium value, with negative $\Delta z$ corresponding to compression.
For each structure characterized by a specific $\Delta z$ value, we Wannierize the same set of 20 DFT bands using the same trial orbitals and $\kv$-point grid of Bloch states, as described in Sec.~\ref{sec:methods}.

Fig.~\ref{fig:characterExchange} shows the results of the Wannierization procedure for $\theta = 1.08^\circ$ TBG for various values of $\Delta z$.
We observe that compression of the layers has the largest impact on the DW orbitals.
For example, in Fig.~\ref{fig:characterExchange} (a), as the amount of compression increases (corresponding to a more negative $\Delta z$ value), the radius of the DW orbitals decreases significantly 
from $\sim 70$ \AA~for uncompressed layers to $\sim 60$ \AA~for a compression of $0.09$ \AA.
The radius of the AB/BA orbitals also decreases slightly, while the radius of the AA orbitals is essentially unchanged.
The decreased radius of the DW orbitals is accompanied by a shift in their centers towards the AA regions, also shown in Fig.~\ref{fig:characterExchange} (a).
In contrast, the AA orbital centers shift slightly away from the AA regions upon compression of the layers.
When the compression reaches $\Delta z = -0.08$ \AA, we observe an exchange of projected orbital character between the upper and lower flat bands near $\Gamma$,
consistent with the exchange of wavefunction character reported in Ref.~\cite{zhu2026wavefunction}.
Such an exchange also occurs when the twist angle decreases to $0.93^\circ$.

Fig.~\ref{fig:characterExchange} (b) shows the on-site and nearest-neighbor Coulomb interactions of the Wannier orbitals for compressed layers.
As the compression increases, the on-site interactions of the DW orbitals increase significantly from $U_{nn}(0)\epsilon \sim 320$ meV at $\Delta z = 0$ to $U_{nn}(0)\epsilon \gtrsim 400$ meV at $\Delta z = -0.09$ \AA, mirroring the trend in the radius and center of the DW orbitals.
The on-site interactions of the AA and AB/BA orbitals also increase slightly as the layers are compressed.
In contrast, the nearest-neighbor interactions remain constant and almost exactly the same for all orbital types.

\begin{figure*}[t]
\centering
\includegraphics[width=1.0\textwidth]{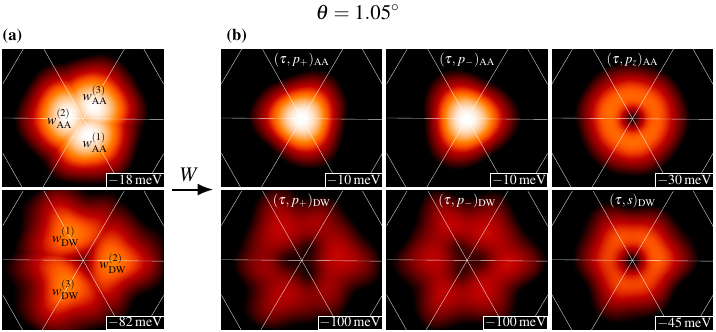}
\caption{%
Transformation of the original Wannier orbitals to the basis in Eq.~\eqref{eq:repnew}.
{\bf (a)} Original Wannier orbitals produced by \textsc{Wannier90}.
{\bf (b)} Symmetrized Wannier orbitals (for details see Appendix). The inset at the bottom of each panel gives the on-site energies of the orbitals relative to $E_{\rm F}$. 
}
\label{fig:symmetry}
\end{figure*}

\section{Discussion}
\subsection{Symmetry and hopping parameters}

Compared to the orbitals in previous tight-binding and many-body models of TBG~\cite{song2022magic,calderon2020interactions,carr2019derivation,po2019faithful}, 
the orbitals produced by the first-principles Wannierization in this work differ slightly in their positions in the moir\'e unit cell.
For example, previous models include $p_\pm$ and $p_z$ orbitals centered at triangular lattice sites and $s$ orbitals at Kagome lattice sites.
In comparison, the orbitals directly produced by \textsc{Wannier90} are systematically offset from the triangular and Kagome lattice sites.
We attribute this feature to differences in the Wannierization procedures.
For example, in the ten-band model of Ref.~\cite{carr2019derivation}, the Wannier orbitals are centered at the triangular and Kagome lattice sites,
whereas no such constraints were imposed in our use of \textsc{Wannier90}.
Furthermore, \textsc{Wannier90} is designed to ``maximally localize'' the total spread of the orbitals 
without explicit consideration of the orbital symmetry.
Nevertheless, we symmetrize the orbitals and tight-binding Hamiltonians to extract the relevant physics and compare with previous tight-binding and many-body models of TBG. A detailed discussion of the symmetrization procedure is provided in Appendix~\ref{sec:symmetrization_appendix}.

\begin{table}[t]
\centering
\setlength{\tabcolsep}{3pt}
\settowidth{\charlabelwidth}{Orbital character}
\setlength{\characterscolwidth}{\dimexpr(\columnwidth-6\tabcolsep-\charlabelwidth)/3\relax}
\begin{tabular}{@{}cccc@{}}
\hline
\hline
\charlabelcell{Orbital character} & \characterscell{$C_3$}            & \characterscell{$C_{2y}$ $(\mathcal{M}_y)$}  & \characterscell{$C_2\mathcal{T}$}  \\
\hline
\charlabelcell{$s$}               & \characterscell{$s$}              & \characterscell{$s$}                         & \characterscell{$s$}               \\
\charlabelcell{$p_z$}             & \characterscell{$p_z$}            & \characterscell{$-p_z$}                      & \characterscell{$p_z$}             \\
\charlabelcell{$p_+$}             & \characterscell{$\omega p_+$}     & \characterscell{$p_-$}                       & \characterscell{$p_-$}             \\
\charlabelcell{$p_-$}             & \characterscell{$\omega^* p_-$}   & \characterscell{$p_+$}                       & \characterscell{$p_+$}             \\
\hline
\hline
\end{tabular}
\caption{
Transformation of the different orbital characters under the $C_3$, $C_{2y}$, and $C_2\mathcal{T}$ symmetries~\cite{po2019faithful}.
We let $\omega = e^{2\pi i / 3}$.
Because $C_2\mathcal{T}$ is antiunitary, its entries are defined only once the phases of the orbitals are chosen (see
Appendix~\ref{sec:symmetrization_appendix} for details).
$C_2\mathcal{T}$ additionally exchanges the AB and BA sites, whereas $C_3$ and $C_{2y}$ leave them fixed.
}
\label{tab:characters}
\end{table}

The symmetrized version of the \textsc{Wannier90} model satisfies all TBG symmetries exactly, and obeys the 
representation matching equation that has been shown to resolve the topological obstruction in TBG~\cite{po2019faithful}.
The symmetrization procedure produces the following set of ten basis orbitals,
\begin{equation}
    \begin{aligned}
        &(\tau, p_z)_{\rm AA} \oplus (\tau, p_\pm)_{\mathrm{AA}} \oplus (\tau, s)_{\rm DW} \oplus (\tau, p_\pm)_{\mathrm{DW}} \oplus (\eta, p_\pm),
    \end{aligned}
    \label{eq:repnew}
\end{equation}
where $\tau$ and $\eta$ denote the centers of the Wannier orbitals and correspond to the triangular and honeycomb lattices, respectively,
and $s$, $p_z$, and $p_\pm$ refer to the atomic-like symmetry of the Wannier orbitals, tabulated in Table~\ref{tab:characters}.
We use the subscript AA (DW) to denote orbitals that are predominantly composed of the original AA (DW) orbitals ($\sim 80$\% to $90$\%).
We refer to this new basis as the ``symmetrized basis.''

Fig.~\ref{fig:symmetry} shows the $\tau$-centered symmetrized orbitals for $\theta = 1.05^\circ$, with the orbitals produced by \textsc{Wannier90} in (a) and the symmetrized orbitals in (b).
We note that the symmetrized orbitals, especially $(\tau, s)_{\rm DW}$, $(\tau, p_z)_{\rm AA}$, and $(\tau, p_\pm)_{\rm DW}$, are 
more delocalized than the orbitals directly obtained by \textsc{Wannier90}.
Furthermore, the symmetrized orbitals do not \emph{perfectly} respect the symmetries outlined in Table~\ref{tab:characters}.
Due to the atomic-scale details of the orbitals, achieving exact symmetry in the orbitals would require a more complicated procedure, 
such as symmetry-adapted Wannierization~\cite{sakuma2013symmetry,pizzi2020wannier90}.

\begin{figure*}[ht]
\centering
\includegraphics[width=1.0\textwidth]{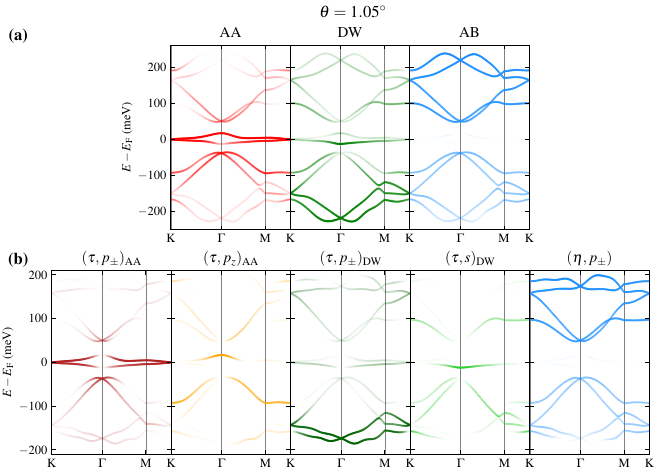}
\caption{%
Projected bands in the ten-band model.
{\bf (a)} Projection onto the original basis: AA, DW, and AB/BA orbitals.
{\bf (b)} Projection onto the symmetrized basis: $(\tau, p_\pm)_{\rm AA}$, $(\tau, p_z)_{\rm AA}$, $(\tau, p_\pm)_{\rm DW}$, $(\tau, s)_{\rm DW}$, and $(\eta, p_\pm)$.
}
\label{fig:projectedBands}
\end{figure*}

In Fig.~\ref{fig:projectedBands} we show the band structures obtained from the original Wannier
orbitals and the symmetrized orbitals, along with the contribution of each type of orbital.  
First, we note that the flat bands and closest remote bands remain nearly unchanged upon symmetrization.
Only the farthest remote bands are affected, and they are $\sim 200$ meV away from the CNP and thus unimportant for the low-energy physics of TBG.
This result highlights the importance of including all ten bands in an accurate, first-principles Wannierization of TBG: 
the most remote bands may not be reproduced accurately under the symmetries of TBG, 
but they act as a buffer such that the low-energy bands of interest can be reproduced with high fidelity.
Additionally, it is evident from this analysis that the dominant contributions to the flat bands come from the 
$(\tau,p_{\pm})_{\rm AA}$ orbitals throughout the mBZ, except in the neighborhood of $\Gamma$, where significant contributions come from $(\tau,p_z)_{\rm AA}$, an AA-related orbital (for electron states) and $(\tau,s)_{\rm DW}$, a DW-related orbital (for hole states). These features are consistent with the topological obstruction of the flat bands~\cite{po2019faithful}.
The DW orbitals, particularly $(\tau,s)_{\rm DW}$, make a significant contribution to the flat bands near $\Gamma$, as well as to the more dispersive bands through the rest of the mBZ. Finally, the $(\eta,p_{\pm})$ AB/BA-related orbitals do not contribute to the flat bands, but rather to the dispersive bands throughout the mBZ. 

The predominance of the $(\tau,p_{\pm})_{\rm DW}$ and $(\eta,p_{\pm})$ orbitals in the dispersive bands is consistent with the observation that hoppings involving these orbitals tend to be the largest in magnitude. 
For example, as shown in Table~\ref{tab:hoppings0}, the hopping between $(\tau,p_{\pm})_{\rm DW}$ and $(\eta,p_{\pm})$ orbitals is the largest of all intracell hoppings, and as shown in Table~\ref{tab:hoppings1}, the largest intercell hoppings involve the $(\tau,p_{\pm})_{\rm DW}$ orbitals. 
The full set of hopping coefficients for angles between $\theta = 0.93^\circ$ and $\theta = 1.05^\circ$ is provided in~\cite{DATA}.
For angles above $\theta = 1.05^\circ$, 
the symmetrized model is unable to accurately reproduce the DFT band structure, indicating that the symmetry breaking during Wannierization was too severe to be undone.

\begin{figure}[t]
\centering
\includegraphics[width=\columnwidth]{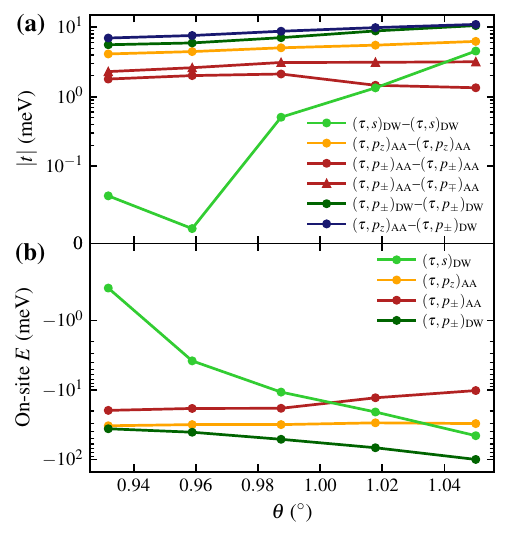}
\caption{%
{\bf (a)} Magnitude of dominant hoppings involving orbitals with significant weight in the flat bands. For $(\tau, p_\pm)_{\rm AA}\text{--}(\tau, p_\mp)_{\rm AA}$, which
contains distinct hoppings in two different $C_3$ orientations (see Appendix~\ref{sec:symmetrization_appendix}), the hoppings
of larger magnitude are shown. {\bf (b)} On-site energy of the $(\tau, s)_{\rm DW}$, $(\tau, p_z)_{\rm AA}$, $(\tau, p_\pm)_{\rm AA}$, and $(\tau, p_\pm)_{\rm DW}$ orbitals.
}
\label{fig:hopping_vs_angle}
\end{figure}

We next discuss the twist-angle dependence of the nearest-neighbor hoppings and on-site energies, shown in Fig.~\ref{fig:hopping_vs_angle}.
The greatest changes involve the $(\tau, s)_{\rm DW}$ orbital.
As the twist angle decreases, the on-site energy of the $(\tau, s)_{\rm DW}$ orbital increases sharply compared to the on-site energy of the $(\tau, p_\pm)_{\rm AA}$ and $(\tau, p_z)_{\rm AA}$ orbitals, mirroring a 
decrease in the hopping between nearest-neighbor $(\tau, s)_{\rm DW}$ orbitals. 
In contrast, the hoppings between other nearest-neighbor orbitals remain relatively constant even as the twist angle decreases.
(We note that although the on-site energy of the $(\tau, p_\pm)_{\rm DW}$ orbitals also increases by $\sim 50$ meV over this range,
those orbitals are predominant in the lowest-energy remote bands.)
The decrease in magnitude of dominant hoppings involving the $(\tau, s)_{\rm DW}$ orbitals as the twist angle decreases from $1.05^\circ$ to $0.93^\circ$ is consistent with the 
rapidly decreasing dispersion of the lower flat band near the $\Gamma$ point, where the $(\tau, s)_{\rm DW}$ orbital dominates. 
Furthermore, between $0.96^\circ$ and $0.93^\circ$, 
the character of the flat bands inverts near $\Gamma$, and during the inversion, the dispersion near $\Gamma$ becomes
very small, mirroring the decrease of the $(\tau, s)_{\rm DW}$ nearest-neighbor hopping to below $0.1$ meV.
This is an interesting finding: as the twist angle decreases, and the moir\'e scale increases, the on-site energy of the DW symmetrized orbitals {\em increases} by around $50$ meV relative to the on-site energies of the $(\tau, p_z)_{\rm AA}$ and $(\tau, p_\pm)_{\rm AA}$ orbitals, while the hopping between $(\tau, s)_{\rm DW}$ orbitals at nearest-neighbor sites within the moir\'e cell, which used to be significant at 5 meV, becomes almost negligible. 

The fact that only the hopping between nearest-neighbor $(\tau, s)_{\rm DW}$ orbitals decreases 
by such a large factor suggests that this hopping plays a major role in determining the dispersion 
of the flat bands, particularly in the lower flat band near the $\Gamma$ point, 
since the $(\tau, s)_{\rm DW}$ orbital dominates the character of those electronic states, 
as shown in Fig.~\ref{fig:projectedBands} (b). 
The absence of this orbital in the upper flat band near $\Gamma$, which is instead dominated by the 
$(\tau, p_z)_{\rm AA}$ orbital whose hoppings remain relatively constant with twist angle, 
further suggests that the hoppings between nearest-neighbor $(\tau, s)_{\rm DW}$ orbitals may be 
responsible for particle-hole asymmetry in superconducting and correlated phases. 
Such an implication is consistent with experimental evidence thus far showing that superconductivity 
is more robust with hole doping at angles slightly above $1^\circ$~\cite{balents2020superconductivity} 
and with electron doping at a smaller angle of $0.93^\circ$~\cite{codecido2019correlated}, 
corresponding to the angles in our model in which the $(\tau, s)_{\rm DW}$ orbital dominates the lower 
and upper flat band, respectively. Determining the role of this hopping at different fillings and in 
the presence of electronic interactions, which have been shown to reshape the bands in both theoretical 
models and experimental measurements~\cite{song2022magic,kim2026ab,xiao2026imaging}, is an important future direction.

\subsection{Coulomb and exchange integrals}

Turning next to a discussion of Coulomb and exchange interactions as described by different theoretical models, Table~\ref{tab:interactions_comparisons} gives a comparison of our results calculated with the symmetrized orbitals to those in an 
eight-band model~\cite{calderon2020interactions} and the heavy fermion model of TBG~\cite{song2022magic}.
To make such comparisons meaningful due to the different number of orbitals in each model, we compare two types of orbitals: 
(I) orbitals with significant contributions to the flat bands, and (II) orbitals with secondary contributions to the flat bands.
For example, in our work, orbitals of type I are the $(\tau, p_\pm)_{\rm AA}$ orbitals while orbitals of type II are the $(\tau, p_z)_{\rm AA}$, $(\tau, s)_{\rm DW}$, and $(\tau, p_\pm)_{\rm DW}$ orbitals.
In the heavy fermion model, orbitals of type I are the $f$ orbitals (which are centered at the triangular lattice sites with $p_\pm$ symmetry, hence similar in character to our $(\tau, p_\pm)_{\rm AA}$ orbitals), while orbitals of type II are the $c$ orbitals.
To compare our values directly with those in the eight-band model, we take $\epsilon = 12$ as used in Ref.~\cite{calderon2020interactions}.
However, as each work assumes a different value of $\epsilon$ and different screening potentials, we compare not only the absolute values of the interaction parameters, 
but also the ratio of each interaction parameter to the on-site Coulomb integral of the type I orbitals in each model, $U_{nn}^{\mathrm{I}}(0)$.

In Table~\ref{tab:interactions_comparisons}, we report comparisons of our Coulomb and exchange interaction values in the symmetrized basis with Refs.~\cite{calderon2020interactions} and~\cite{song2022magic}.
We show values at two twist angles, $1.05^\circ$ and $0.96^\circ$ (the magic angle in our DFT calculations), and find that they are similar.
As such, we primarily compare the values calculated at $1.05^\circ$.
With $\epsilon = 12$, we find a value of $U_{nn}^{\mathrm{I}}(0) = 39$ meV at $\theta = 1.05^\circ$, in reasonable agreement with $U_{nn}^{\mathrm{I}}(0) \sim 42$ meV for the corresponding orbitals in Ref.~\cite{calderon2020interactions} (Table~\ref{tab:interactions_comparisons}, first row).
The on-site interactions for our type II orbitals, $U_{nn}^{\mathrm{II}}(0)$, are also similar to those in Ref.~\cite{calderon2020interactions}, both in their absolute values and ratios to the type I on-site $U$ in each model (Table~\ref{tab:interactions_comparisons}, second row).
This result may be expected due to the similarity in the character and positions of the orbitals in each model.

Comparison of the absolute value of on-site interaction terms with the heavy fermion model is more difficult due to its use of a more complex dielectric screening potential.
Furthermore, it should be noted that the on-site interactions of the $f$ orbitals in the heavy fermion model are sensitive to the parameters chosen for the Bistritzer--MacDonald model~\cite{Bistritzer2011}, such as $v_{\rm F}$ and $w_0/w_1$, and can vary from approximately $37$ to $70$ meV for reasonable choices of parameters~\cite{song2022magic}.
However, taking a representative value of $U_{nn}^{\mathrm{I}}(0) \sim 58$ meV for the $f$ orbitals and $U_{nn}^{\mathrm{II}}(0) \sim 48$ meV for the $c$ orbitals in the heavy fermion model, we find that the ratio of $c$ to $f$ on-site interactions is larger than the ratio of type II to type I on-site interactions in the symmetrized Wannier model, as shown in the second row of Table~\ref{tab:interactions_comparisons}.

\begin{table}[t]
\centering
\setlength{\nnlabelcolwidth}{0.40\columnwidth}
\setlength{\nnvaluecolwidth}{\dimexpr(\columnwidth-\nnlabelcolwidth-8\tabcolsep)/4\relax}
\begin{tabular}{@{}c|cc|c|c@{}}
\hline
\hline
& \multicolumn{2}{c|}{This work}   & \multirow{2}{*}{\nnvaluecell{\cite{calderon2020interactions}}}  & \multirow{2}{*}{\nnvaluecell{\cite{song2022magic}}} \\
\cline{2-3}
& \nnvaluecell{$0.96^\circ$}   & \nnvaluecell{$1.05^\circ$}   & & \\
\hline
\nnlabelcell{$U_{nn}^{\mathrm{I}}(0)$ (meV)}                & \nnvaluecell{$37$}        & \nnvaluecell{$39$}        & \nnvaluecell{$42$}                        & \nnvaluecell{$58$} \\
\hline
\nnlabelcell{$U_{nn}^{\mathrm{II}}(0)$ (meV)}            & \nnvaluecell{$23$}       & \nnvaluecell{$21$}       & \nnvaluecell{$21$}                       & \nnvaluecell{$48$}  \\
\nnlabelcell{$U_{nn}^{\mathrm{II}}(0) / U_{nn}^{\mathrm{I}}(0)$}            & \nnvaluecell{$0.62$}       & \nnvaluecell{$0.54$}       & \nnvaluecell{$0.5$}                       & \nnvaluecell{$0.83$}  \\
\hline
\nnlabelcell{$U_{nn}(\vert \R \vert = \lambda)$ (meV)}               & \nnvaluecell{$8.8$}        & \nnvaluecell{$9.6$}        & \nnvaluecell{$9.4$}                        & \nnvaluecell{$2.3$} \\
\nnlabelcell{$U_{nn}(\vert \R \vert = \lambda) / U_{nn}^{\mathrm{I}}(0)$}            & \nnvaluecell{$0.24$}       & \nnvaluecell{$0.24$}       & \nnvaluecell{$0.22$}                       & \nnvaluecell{$0.04$}  \\
\hline
\nnlabelcell{$J_{nm}$ (meV)}            & \nnvaluecell{$\leq 6.7$}       & \nnvaluecell{$\leq 6.2$}       & \nnvaluecell{$\leq 3$}                       & \nnvaluecell{$16$}  \\
\nnlabelcell{$J_{nm} / U_{nn}^{\mathrm{I}}(0)$}            & \nnvaluecell{$\leq 0.18$}       & \nnvaluecell{$\leq 0.16$}       & \nnvaluecell{$\leq 0.07$}                       & \nnvaluecell{$0.28$}  \\
\hline
\hline
\end{tabular}
\caption{
Comparison of Coulomb and exchange interaction parameters in our ten-band model (taking $\epsilon = 12$), an eight-band model~\cite{calderon2020interactions}, and a heavy fermion model~\cite{song2022magic}, all with $\theta = 1.05^\circ$; for our model, values at $\theta = 0.96^\circ$ are also shown.
$U_{nn}^{\mathrm{I}}(0)$ denotes the on-site interactions of the type I orbitals (those predominant in the flat bands), and $U_{nn}^{\mathrm{II}}(0)$ denotes the on-site interactions of the type II orbitals (those predominant in the remote bands with secondary contribution to the flat bands).
}
\label{tab:interactions_comparisons}
\end{table}

The third row of Table~\ref{tab:interactions_comparisons} compares the nearest-neighbor Coulomb interactions in these three models.
Because the nearest-neighbor interactions in our model and in Ref.~\cite{calderon2020interactions} are nearly independent of orbital type, comparisons can be made without classifying the orbitals.
The nearest-neighbor interactions in our model and in Ref.~\cite{calderon2020interactions} are quite close, while the nearest-neighbor interactions between $f$ orbitals in the heavy fermion model are significantly smaller, suggesting that the $f$ orbitals in the heavy fermion construction are more localized.

The last row of Table~\ref{tab:interactions_comparisons} compares the exchange interaction parameters between type I and type II orbitals in our model with those in the other two models.
Since the exchange interactions in our model and the eight-band model span a range of values depending on the specific orbitals involved, we compare the largest exchange interaction in the eight-band model and in our model, which is between the $(\tau, p_\pm)_{\mathrm{AA}}$ and $(\tau, p_z)_{\rm AA}$ orbitals, against the exchange interaction between the $f$ and $c$ orbitals in the heavy fermion model.
We find that the ratio of the exchange interaction to the type I on-site $U$ in our model is $\sim 0.16$, larger than the ratio of $\sim 0.07$ in the eight-band model and smaller than the ratio of $\sim 0.28$ for the exchange interaction between the $f$ and $c$ orbitals in the heavy fermion model.
Note that the exchange value we report is between orbitals in the \textit{symmetrized} basis, whereas those reported in Fig.~\ref{fig:U} are between orbitals in the original \textsc{Wannier90} basis.

\section{Conclusions}

In this work we have presented tight-binding models with real-space Wannier orbitals for TBG over a range of twist angles including the magic angle, derived entirely from first principles.
By Wannierizing the 20 DFT bands closest to the Fermi level, we obtain a ten-band model for each of the two valleys.
Using the real-space orbitals, we compute on-site and nearest-neighbor Coulomb interaction parameters, finding that the on-site interaction values of the AA orbitals are largest and decrease with decreasing twist angle, while the nearest-neighbor interaction values are approximately one quarter of the on-site interaction values and nearly independent of orbital type.
We study the effect of increased interlayer interactions on the Wannier states by compressing the layers of TBG, finding that the DW orbitals are most affected by compression and confirming that a critical interlayer interaction strength exists at which an exchange of character occurs between the upper and lower flat bands.
Finally, we symmetrize the tight-binding models and orbitals and observe hoppings between certain flat-band orbitals to exhibit a strong dependence on the twist angle within the magic range.

The detailed comparison of various interactions leads us to expect 
that in a Hubbard-like model based on the Wannier states the
nearest-neighbor Coulomb interactions will play a larger role 
than they would in a model based on the heavy fermion approach.
However, the on-site interactions in our model are still by far the dominant energy scale, and thus we expect them to play the largest role in many-body models.
Hartree--Fock calculations will be an important future step to draw further comparisons between our models and similar works~\cite{song2022magic,calderon2020interactions,kim2026ab} and to determine the importance of the different interactions for the many-body physics of TBG.
By using the orbitals and interactions in our ten-band model as input for such calculations, we open the possibility of modeling the many-body physics of TBG entirely from first principles to make more rigorous comparisons with experimental observations of the many-body ground state of TBG.

\section{Acknowledgments}
The authors acknowledge support from the Simons Foundation award No.~896626.
A.Z.~is supported by the U.S.~Department of Energy, Office of Science, Office of Advanced Scientific Computing Research, Department of Energy Computational Science Graduate Fellowship under Award Number DE-SC0025528.
D.B.~acknowledges support from the NTU Startup Grant (Award Number 025661-00003).
M.M.A.E.~acknowledges support from the Ministry of Education, Singapore (Research Centre of Excellence award to the Institute for Functional Intelligent Materials, I-FIM, project No. EDUNC-33-18-279-V12) and the National Research Foundation, Singapore, under its AI Singapore Programme (AISG Award No: AISG3-RP-2022-028).
E.~M.~acknowledges the hospitality of the Harvard Physics Department, where part of this work was done. 
Computing resources were provided by the Harvard University FAS Division of Science Research Computing Group and the Texas Advanced Computing Center (TACC) at The University of Texas at Austin.

\section{Data Availability}
Relaxed atomic structures, DFT band structures, and symmetrized tight-binding model parameters are publicly available at the GitHub link in~\cite{DATA}.

\section{Use of AI Tools}
The authors used Claude Code (Opus 5) to assist with the implementation of the symmetrization procedure of the tight-binding Hamiltonians and orbitals described in the Appendix. All AI-generated code was independently verified by the authors who take full responsibility for the accuracy of the content of this manuscript.

\appendix
\setcounter{figure}{0}
\setcounter{table}{0}
\renewcommand{\thefigure}{A\arabic{figure}}
\renewcommand{\thetable}{A\arabic{table}}
\renewcommand{\theHfigure}{A\arabic{figure}}
\renewcommand{\theHtable}{A\arabic{table}}
\section{Symmetrization of the \textsc{Wannier90} tight-binding model
\label{sec:symmetrization_appendix}
}

\subsection{Symmetry operations and representation matrices}

The tight-binding Hamiltonians produced by \textsc{Wannier90} do not exactly obey the lattice symmetries of TBG.
This is an artifact of the Wannierization rather than a property of the underlying physics, as it was previously shown that the relaxed structure of TBG obeys a universal $D_6$ symmetry~\cite{ezzi2026universal}.
Here we show that the tight-binding models can be projected onto a subspace of Hamiltonians that exactly obey these symmetries.
We note that an alternative route is to impose the symmetry during the Wannierization itself, 
constraining the spread minimization so that the orbitals come out symmetry-adapted by 
construction~\cite{sakuma2013symmetry,pizzi2020wannier90}. However, symmetry-adapted Wannierization requires symmetry information about the Bloch states 
that the interface between \textsc{Siesta} and \textsc{Wannier90} does not currently provide.

Working within a single valley, the relevant operations are those that do not exchange the two valleys: $C_3$, a threefold rotation about the $z$-axis
defined to be perpendicular to the plane of the bilayer; $C_{2y}$, a twofold rotation about the $y$-axis defined to be
parallel to the plane of the bilayer (sometimes referred to as $\mathcal{M}_y$ in purely 2D models); and $C_2\mathcal{T}$, a twofold rotation about the $z$-axis followed by time-reversal symmetry (complex conjugation).
The operations $C_3$, $C_{2y}$, $C_2\mathcal{T}$, and their compositions form a group $G$ with $\vert G \vert = 12$ elements acting within a single valley.
Defining $D(g)$ to be the $10 \times 10$ representation matrix of $g \in G$ in the basis of the 10 Wannier orbitals $\vert w_n \rangle$ of a single valley, we have
\begin{equation}
    g\vert w_n \rangle = \sum_m D_{mn}(g) \vert w_m \rangle,
    \label{eq:repmatrix}
\end{equation}
with $D(g)^{\dagger}D(g) = I$ for every $g \in G$ (both unitary and antiunitary). Furthermore, in a composition $g_i g_j$ in which $g_j$ acts first, $D(g_i g_j) = D(g_i) D(g_j)$ if $g_i$ is unitary
and $D(g_i g_j) = D(g_i) D(g_j)^*$ if $g_i$ is antiunitary, in either case irrespective of whether $g_j$ is unitary or antiunitary.
Let $g\R$ denote the image of a moir\'e lattice vector $\R$ under $g$, and $g\kv$ the image of a momentum vector $\kv$, which for antiunitary $g$ includes the reversal $\kv \to -\kv$ due to $\mathcal{T}$.
A matrix of hoppings $H(\R)$ from orbitals in the unit cell translated by $\R$ to orbitals in the home cell must then satisfy
\begin{equation}
    H(g\R) = D(g) H(\R) D(g)^{\dagger},
    \label{eq:symcond}
\end{equation}
with complex conjugation applied to $H(\R)$ if $g$ is antiunitary.
Equivalently, in terms of the Bloch Hamiltonian $H(\kv) = \sum_{\R} e^{i \kv \cdot \R} H(\R)$,
\begin{equation}
    H(g\kv) = D(g) H(\kv) D(g)^{\dagger},
    \label{eq:symcondk}
\end{equation}
again with $H(\kv)$ conjugated if $g$ is antiunitary.

\subsection{Determining the representation matrices}

The matrices $D(g)$ are not known in advance.
\textsc{Wannier90} returns the orbitals in a basis that minimizes the total spread of the orbitals, 
so the way each operation permutes and mixes the basis orbitals must be measured from the model itself.
We do this by fitting $D(g)$ to satisfy Eq.~\eqref{eq:symcondk} using the \textsc{Wannier90} Hamiltonian $H_0$, one generator at a time.
Only the matrices of the three generators, $D(C_3)$, $D(C_{2y})$, and $D(C_2\mathcal{T})$, are fitted; the representations of the remaining nine elements of $G$ 
are obtained from them by composition.
For a fixed $g$, define
\begin{equation}
    B_g(\kv) =
    \begin{cases}
        H_0(g^{-1}\kv), & g \;\text{unitary}, \\
        H_0(g^{-1}\kv)^*, & g \;\text{antiunitary},
    \end{cases}
\end{equation}
so that Eq.~\eqref{eq:symcondk} becomes $D(g) B_g(\kv) D^{\dagger}(g) = H_0(\kv)$ at every $\kv$.
Since no $D(g)$ satisfies this equality exactly due to the broken symmetries, we take $D(g)$ to be the unitary matrix that minimizes the difference,
\begin{equation}
    D(g) = \underset{D^{\dagger}D = I}{\mathrm{arg\,min}} \; \sum_{\kv} \big\lVert H_0(\kv) - D B_g(\kv) D^{\dagger} \big\rVert^2 ,
    \label{eq:procrustes}
\end{equation}
where where $\lVert X\rVert$ is the Frobenius norm of $X$, $\lVert X\rVert^2 = \sum_{ij}\vert X_{ij}\vert^2$, and the sum runs over a uniform $12 \times 12$ grid of $\kv$ points in the mBZ.
Equation~\eqref{eq:procrustes} is a two-sided orthogonal Procrustes problem~\cite{schonemann1968two}, which has no closed-form solution due to the sum over $\mathbf{k}$. We thus solve it by iterating
\begin{equation}
    D^{(i+1)} = \mathcal{P}\left[ \mathcal{U}\left( \sum_{\kv} H_0(\kv) \, D^{(i)} \, B_g(\kv)^{\dagger} \right) \right]
    \label{eq:procrustesiter}
\end{equation}
from an initial guess $D^{(0)}$, where the superscript labels the iteration, until the maximum element of
$\vert D^{(i+1)} - D^{(i)}\vert$ becomes less than $10^{-11}$ or $500$ iterations are reached. 
Several initial guesses $D^{(0)}$ are used, and the one that achieves the minimum of Eq.~\eqref{eq:procrustes} is taken as $D(g)$.

In each iteration, we first form the matrix $X = \sum_{\kv} H_0(\kv) \, D^{(i)} \, B_g(\kv)^{\dagger}$ appearing in Eq.~\eqref{eq:procrustesiter} and compute
$Y = \mathcal{U}(X) = UV^{\dagger}$, the unitary matrix nearest to $X$, where $U$ and $V$ come from the singular value decomposition $X = U \Sigma V^{\dagger}$~\cite{fan1955metric}.
Each iteration thereby holds $D^{\dagger}$ on the right-hand side of Eq.~\eqref{eq:procrustes} fixed at its current value, which leaves a one-sided Procrustes problem that $Y=UV^\dagger$ solves exactly~\cite{schonemann1966generalized}.
We then compute $\mathcal{P}(Y)$, the projection of $Y$ onto the subspace of matrices that possess the structure of a genuine representation of $g$.
$\mathcal{P}(Y)$ first enforces $Y$ to only connect orbitals that can be transformed into each other by $g$ based on their positions in the moir\'e cell.
Letting $\vert w_\tau \rangle$ denote either an AA orbital $\vert w_{\rm AA} \rangle$ or DW orbital $\vert w_{\rm DW} \rangle$, and letting
$\vert w_{\rm AB} \rangle$ and $\vert w_{\rm BA} \rangle$ denote the AB and BA orbitals, respectively,
\begin{itemize}
    \item $D(C_3)$ can only have nonzero elements connecting $\vert w_\tau \rangle$ to other $\vert w_\tau \rangle$ (including itself), 
    $\vert w_{\rm AB} \rangle$ to $\vert w_{\rm AB} \rangle$, and $\vert w_{\rm BA} \rangle$ to $\vert w_{\rm BA} \rangle$;
    \item $D(C_{2y})$ can only have nonzero elements connecting $\vert w_\tau \rangle$ to $\vert w_\tau \rangle$, 
    $\vert w_{\rm AB} \rangle$ to $\vert w_{\rm AB} \rangle$, and $\vert w_{\rm BA} \rangle$ to $\vert w_{\rm BA} \rangle$;
    \item $D(C_2\mathcal{T})$ can only have nonzero elements connecting $\vert w_\tau \rangle$ to $\vert w_\tau \rangle$, 
    $\vert w_{\rm AB} \rangle$ to $\vert w_{\rm BA} \rangle$, and $\vert w_{\rm BA} \rangle$ to $\vert w_{\rm AB} \rangle$.
\end{itemize}
As a result, $D(C_3)$ and $D(C_{2y})$ have 3 diagonal blocks corresponding to the orbitals at each of the three high-symmetry stacking centers (AA, AB, and BA), while $D(C_2\mathcal{T})$ has 2 diagonal blocks corresponding to the AA and AB/BA sites.

$\mathcal{P}(Y)$ imposes further structure on $Y$ for each $g$. Since $D(C_3)^3 = I$, the eigenvalues $\lambda_b$ of $D(C_3) = P\mathrm{diag}(\lambda_b)P^{-1}$ 
must be the cube roots of unity $\omega^{m_b}$ for $m_b \in \{0, \pm 1\}$ and $\omega = e^{2\pi i/3}$. $\mathcal{P}$ enforces this structure by rounding each
$\lambda_b$ to the nearest cube root of unity and rebuilding $D(C_3)$ as $P\mathrm{diag}(\omega^{m_b})P^{-1}$ with $P$ being a unitary matrix.
Furthermore, within each diagonal block of $D(C_3)$, the number of eigenvalues with $m_b = +1$ must equal the number of eigenvalues with $m_b = -1$ as
required by $C_{2y}$; a $C_{2y}$ rotation flips the eigenvalue of an orbital under $C_3$ and thus must map an orbital with eigenvalue $\omega^{m_b}$ 
to an orbital with eigenvalue $\omega^{-m_b}$.
Solutions that do not satisfy this constraint are discarded. Similar structural constraints can be derived for $D(C_{2y})$ and $D(C_2\mathcal{T})$:
\begin{itemize}
    \item $D(C_{2y})^2 = I$ and $D(C_{2y})^{\dagger}D(C_{2y}) = I$, so $D(C_{2y}) = D(C_{2y})^{\dagger}$ and $\mathcal{P}$ maps
    $Y$ to $\frac{1}{2}\left(Y + Y^{\dagger}\right)$.
    \item $D(C_2\mathcal{T})D(C_2\mathcal{T})^{*} = I$ and $D(C_2\mathcal{T})^{\dagger}D(C_2\mathcal{T}) = I$, so
    $D(C_2\mathcal{T})^{*} = D(C_2\mathcal{T})^{\dagger} = \left(D(C_2\mathcal{T})^{*}\right)^{\top} \implies D(C_2\mathcal{T}) = D(C_2\mathcal{T})^{\top}$ and
    $\mathcal{P}$ maps $Y$ to $\frac{1}{2}\left(Y + Y^{\top}\right)$.
\end{itemize}
Since $C_2\mathcal{T}$ commutes with $C_{2y}$, which for an antiunitary $D(C_2\mathcal{T})$ requires $D(C_2\mathcal{T}) = D(C_{2y}) D(C_2\mathcal{T}) D(C_{2y})^{\top}$ rather than ordinary matrix commutation,
$\mathcal{P}$ further maps $D(C_2\mathcal{T})$ to $\frac{1}{2}\left(D(C_2\mathcal{T}) + D(C_{2y}) D(C_2\mathcal{T}) D(C_{2y})^{\top}\right)$.
Unitarity is then restored by a second application of $\mathcal{U}$.

The representations of $C_3$, $C_{2y}$, and $C_2\mathcal{T}$ are fitted sequentially. 
$D(C_3)$ is fitted first; its eigenvectors define a unitary matrix $Q$, and the \textsc{Wannier90} Hamiltonian $H_0$ is rotated into this eigenbasis to form $H_{C_3} = Q^{\dagger} H_0 Q$.
In that basis the representation of $C_3$ becomes diagonal: $Q^{\dagger} D(C_3) Q = \mathrm{diag}(\omega^{m_b})$.
$D(C_{2y})$ and $D(C_2\mathcal{T})$ are then both fitted to $H_{C_3}$, in that same basis, with $D(C_2\mathcal{T})$ fitted last so that it can also be required to commute with the $D(C_{2y})$ just found.
With all representations fitted, the tight-binding Hamiltonian produced by \textsc{Wannier90} $H_0$ 
is projected onto a Hamiltonian $H_{\rm proj}$ that obeys the symmetries of TBG by averaging over the group $G$~\cite{gresch2018automated}:
\begin{equation}
    H_{\rm proj}(\R) = \frac{1}{|G|} \sum_{g \in G} D(g) H_0(g^{-1}\R) D(g)^{\dagger},
    \label{eq:groupaverage}
\end{equation}
again with complex conjugation applied to $H_0$ if $g$ is antiunitary.


\subsection{Relation to other ten-band models}

The construction of $H_{\rm proj}$ also reveals the relation between the basis of the tight-binding model produced by
\textsc{Wannier90} and the basis of previously proposed ten-band models, such as that in Ref.~\cite{po2019faithful}. As shown below, the basis of $H_{\rm proj}$ can be transformed to satisfy the representation-matching equation,
\begin{equation}
    \begin{aligned}
        &(\tau, p_z)_{\rm AA} \oplus (\tau, p_\pm)_{\mathrm{AA}} \oplus (\tau, s)_{\rm DW} \oplus (\tau, p_\pm)_{\mathrm{DW}} \oplus (\eta, p_\pm)_0 \\
        &\qquad \overset{\mathrm{rep.}}{=} (\eta, p_\pm)_1 \oplus (\mathrm{target}) \oplus (\eta, p_\pm)_2,
    \end{aligned}
    \label{eq:repnew_appendix}
\end{equation}
thereby respecting the lattice symmetries of TBG and resolving the topological obstruction. 
Here, $\tau$ and $\eta$ denote the centers of the Wannier orbitals and correspond to the triangular (AA-stacking) and honeycomb (AB and BA) lattices, respectively.
The characters $s$, $p_z$, and $p_\pm$ refer to the atomic-like symmetry of the Wannier orbitals, with ``target'' referring to the representations required for the flat bands.
(The subscripts $0$ to $2$ are added following the convention in Ref.~\cite{po2019faithful} to denote that the orbitals have different physical degrees of freedom.)
Table~\ref{tab:characters} shows the transformation of the $s$, $p_z$, and $p_\pm$ orbitals under these symmetries.

As a result of the projection previously described, the basis of $H_{\rm proj}$ possesses the $C_3$ eigenvalues in Table~\ref{tab:characters}. 
To make the other representation matrices, $D(C_{2y})$ and $D(C_2\mathcal{T})$, match those of Table~\ref{tab:characters}, 
we apply a unitary transformation $\Phi$ to $H_{\rm proj}$,
\begin{equation}
    H_{\rm proj}(\R) \longrightarrow \Phi^{\dagger} H_{\rm proj}(\R)\, \Phi .
    \label{eq:pobasis}
\end{equation}
Under $\Phi$, the representation matrices transform as $D(g) \to \Phi^{\dagger} D(g)\, \Phi$ for unitary $g$ and as $D(g) \to \Phi^{\dagger} D(g) \Phi^{*}$ for antiunitary $g$.
$\Phi$ must preserve the representation $D(C_3) = \mathrm{diag}(\omega^{m_b})$, requiring $\Phi^{\dagger} D(C_3) \Phi = D(C_3)$, i.e.,
\begin{equation}
    \Phi_{ab} \left( \omega^{m_a} - \omega^{m_b} \right) = 0 ,
    \label{eq:phicommute}
\end{equation}
so that $\Phi_{ab} = 0$ unless $m_a = m_b$.
$\Phi$ must also not mix orbitals belonging to different stacking centers.
Together, these requirements render $\Phi$ block diagonal in stacking site (AA, AB, or BA) and $C_3$ eigenvalue (1, $\omega$, or $\omega^*$).

To illustrate how $\Phi$ is determined, we consider the six $\tau$-centered orbitals which contain two orbitals per $C_3$ eigenvalue (Eq.~\eqref{eq:repnew_appendix}).
Thus, the block of $\Phi$ acting on the $\tau$ site has the following block-diagonal form,
\begin{equation}
    \Phi_{\tau} = \Phi_0 \oplus \Phi_+ \oplus \Phi_- ,
    \label{eq:phitau}
\end{equation}
where each $\Phi_m$ is a $2 \times 2$ unitary matrix acting within the pair of orbitals with $C_3$ eigenvalue $\omega^m$.
To satisfy Table~\ref{tab:characters}, $\Phi_0$ must satisfy
\begin{equation}
    \Phi_0^{\dagger} \left[D(C_{2y})\right]_0 \Phi_0 = \begin{pmatrix} -1 & 0 \\ 0 & 1 \end{pmatrix},
    \qquad
    \Phi_0^{\dagger} \left[D(C_2\mathcal{T})\right]_0 \Phi_0^{*} = I,
    \label{eq:phi0req}
\end{equation}
where $\left[D(C_{2y})\right]_0$ and $\left[D(C_2\mathcal{T})\right]_0$ denote the blocks of the two representation 
matrices acting on the pair of orbitals with $C_3$ eigenvalue $1$. $\Phi_0$ can thus be determined by diagonalizing
$\left[D(C_{2y})\right]_0$ to find the eigenbasis $U_0$, which determines $\Phi_0$ up to phase on each eigenstate:
\begin{equation}
    \Phi_0 = U_0 \, \mathrm{diag}\!\left(e^{i\varphi_1}, e^{i\varphi_2}\right).
    \label{eq:phi0form}
\end{equation}
$\left[D(C_2\mathcal{T})\right]_0$ is then rotated into the eigenbasis $U_0$, resulting in a unitary diagonal matrix
as $D(C_2\mathcal{T})$ is unitary and $C_2\mathcal{T}$ commutes with $C_{2y}$:
\begin{equation}
    U_0^{\dagger} \left[D(C_2\mathcal{T})\right]_0 U_0^{*}
    = \mathrm{diag}\!\left(e^{i\alpha_1}, e^{i\alpha_2}\right).
    \label{eq:phi0alpha}
\end{equation}
So by Eq.~\eqref{eq:phi0req}, $e^{i(\alpha_b - 2\varphi_b)} = 1 \implies \varphi_b = \alpha_b/2$, which determines $\Phi_0$.

$\Phi_+$ and $\Phi_-$ can be determined in a similar manner.
Requiring that $C_{2y}$ and $C_2\mathcal{T}$ exchange $p_+$ and $p_-$ forces the block of 
$D(C_{2y})$ and $D(C_2\mathcal{T})$ involving those orbitals to 
take the form $\left(\begin{smallmatrix} 0 & I \\ I & 0 \end{smallmatrix}\right)$, 
with the two blocks being labeled by the $C_3$ eigenvalues $\omega$ and $\omega^*$.
Defining $[\,\cdot\,]_{-+}$ as the block of a matrix coupling the $p_+$ and $p_-$ orbitals at the $\tau$ site,
this requirement results in
\begin{equation}
    \Phi_-^{\dagger} \left[D(C_{2y})\right]_{-+} \Phi_+ = I,
    \qquad
    \Phi_-^{\dagger} \left[D(C_2\mathcal{T})\right]_{-+} \Phi_+^{*} = I
    \label{eq:phireq}
\end{equation}
Eliminating $\Phi_-$ between these equations leaves the single condition
\begin{equation}
    \Phi_+ \Phi_+^{\top} = S,
    \qquad
    S = \left[D(C_{2y})\right]_{-+}^{-1} \left[D(C_2\mathcal{T})\right]_{-+} .
    \label{eq:takagi}
\end{equation}

Since $S$ is a symmetric, unitary matrix, it can be written as $S = V \mathrm{diag}(e^{i\beta_b}) V^{\top}$~\cite{horn2013matrix,ikramov2012takagi}, so
\begin{equation}
    \Phi_+ = V \mathrm{diag}\left(e^{i\beta_b/2}\right) V^{\top},
\end{equation}
where $V$ is a real orthogonal matrix. We then have $\Phi_- = \left[D(C_{2y})\right]_{-+} \Phi_+$. Note that Eq.~\eqref{eq:takagi} does not fix $\Phi_+$ completely.
If $\Phi_+$ is one solution, then so is $\Phi_+ O$ for any real orthogonal matrix $O$:
\begin{equation}
    \left(\Phi_+ O\right)\left(\Phi_+ O\right)^{\top}
    = \Phi_+ O O^{\top} \Phi_+^{\top} = S.
\end{equation}
We choose to tune $O$ to maximize the difference in the spread between the 
$(\tau, p_\pm)_{\rm AA}$ and $(\tau, p_\pm)_{\rm DW}$ orbitals and take
the final value of $\Phi_+$ to be $\Phi_+ O$.

\begin{figure}[t]
\centering
\includegraphics[width=\columnwidth]{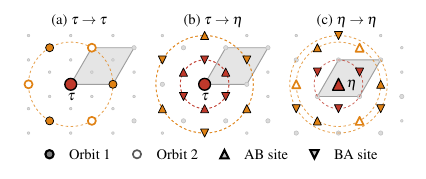}
\caption{%
Orbital positions corresponding to the on-site energies and hoppings listed in Tables~\ref{tab:hoppings0} (red) and~\ref{tab:hoppings1} (orange).
{\bf (a)} Hoppings from an orbital $w_a$ centered at $\tau$ to an orbital $w_b$ centered at $\tau$. 
{\bf (b)} Hoppings from an orbital centered at $\tau$ to an orbital centered at $\eta$.
{\bf (c)} Hoppings from an orbital centered at $\eta$ to an orbital centered at $\eta$.
Dashed circles mark hoppings that are equidistant from the central orbital in each panel.
Filled and empty markers denote two distinct sets of equidistant $C_3$-related neighbors
whose hopping magnitudes can differ.
The shaded parallelograms indicate the moir\'e cell, with lower left corner at $\tau$.
}
\label{fig:shellmap}
\end{figure}

\begin{table}[t]\centering\scriptsize
\begin{tabular}{l|rrrrrrrrrr}
\hline\hline
 & $p_z$ & $p_+^{\rm AA}$ & $p_-^{\rm AA}$ & $s$ & $p_+^{\rm DW}$ & $p_-^{\rm DW}$ & $p_+^{\rm AB}$ & $p_-^{\rm AB}$ & $p_+^{\rm BA}$ & $p_-^{\rm BA}$ \\
\hline
$p_z$ & $-30.40$ & $-0-$ & $-0-$ & $-0-$ & $-0-$ & $-0-$ & 25.10 & 25.10 & 25.10 & 25.10 \\
\hline
$p_+^{\rm AA}$ &  & $-10.17$ & $-0-$ & $-0-$ & 12.70 & $-0-$ & 9.65 & 17.86 & 9.65 & 17.86 \\
\hline
$p_-^{\rm AA}$ &  &  & $-10.17$ & $-0-$ & $-0-$ & 12.70 & 17.86 & 9.65 & 17.86 & 9.65 \\
\hline
$s$ &  &  &  & $-45.39$ & $-0-$ & $-0-$ & 26.39 & 26.39 & 26.39 & 26.39 \\
\hline
$p_+^{\rm DW}$ &  &  &  &  & $-99.76$ & $-0-$ & 11.36 & 41.76 & 11.36 & 41.76 \\
\hline
$p_-^{\rm DW}$ &  &  &  &  &  & $-99.76$ & 41.76 & 11.36 & 41.76 & 11.36 \\
\hline
$p_+^{\rm AB}$ &  &  &  &  &  &  & 74.24 & $-0-$ & 23.74 & 15.62 \\
\hline
$p_-^{\rm AB}$ &  &  &  &  &  &  &  & 74.24 & 15.62 & 23.74 \\
\hline
$p_+^{\rm BA}$ &  &  &  &  &  &  &  &  & 74.24 & $-0-$ \\
\hline
$p_-^{\rm BA}$ &  &  &  &  &  &  &  &  &  & 74.24 \\
\hline\hline
\end{tabular}
\caption{
    On-site energies and magnitude of hoppings (in meV) between orbitals at positions denoted
    by the red markers in Fig.~\ref{fig:shellmap}, for $\theta = 1.05^\circ$. 
    ``$-0-$'' denotes that the hopping is forbidden by symmetry.
    \label{tab:hoppings0}
    }
\end{table}

\begin{table}[t]\centering\scriptsize
\begin{tabular}{l|rrrrrrrrrr}
\hline\hline
 & $p_z$ & $p_+^{\rm AA}$ & $p_-^{\rm AA}$ & $s$ & $p_+^{\rm DW}$ & $p_-^{\rm DW}$ & $p_+^{\rm AB}$ & $p_-^{\rm AB}$ & $p_+^{\rm BA}$ & $p_-^{\rm BA}$ \\
\hline
\multirow{2}{*}{$p_z$} & 6.25 & 2.31 & 4.23 & 2.42 & 10.98 & 2.93 & 3.07 & 3.07 & 3.07 & 3.07 \\
 & & 4.23 & 2.31 & & 2.93 & 10.98 & & & & \\
\hline
\multirow{2}{*}{$p_+^{\rm AA}$} & & 1.34 & 3.18 & 1.64 & 5.76 & 2.00 & 1.66 & 1.96 & 1.66 & 1.96 \\
 & & & 0.86 & 1.76 & 0.42 & 4.24 & & & & \\
\hline
\multirow{2}{*}{$p_-^{\rm AA}$} & & & 1.34 & 1.76 & 4.24 & 0.42 & 1.96 & 1.66 & 1.96 & 1.66 \\
 & & & & 1.64 & 2.00 & 5.76 & & & & \\
\hline
\multirow{2}{*}{$s$} & & & & 4.54 & 3.86 & 6.02 & 2.94 & 2.94 & 2.94 & 2.94 \\
 & & & & & 6.02 & 3.86 & & & & \\
\hline
\multirow{2}{*}{$p_+^{\rm DW}$} & & & & & 10.56 & 5.35 & 5.41 & 2.24 & 5.41 & 2.24 \\
 & & & & & & 7.02 & & & & \\
\hline
$p_-^{\rm DW}$ & & & & & & 10.56 & 2.24 & 5.41 & 2.24 & 5.41 \\
\hline
\multirow{2}{*}{$p_+^{\rm AB}$} & & & & & & & 2.90 & 4.25 & 2.04 & 3.25 \\
 & & & & & & & & 2.94 & & \\
\hline
$p_-^{\rm AB}$ & & & & & & & & 2.90 & 3.25 & 2.04 \\
\hline
\multirow{2}{*}{$p_+^{\rm BA}$} & & & & & & & & & 2.90 & 4.25 \\
 & & & & & & & & & & 2.94 \\
\hline
$p_-^{\rm BA}$ & & & & & & & & & & 2.90 \\
\hline\hline
\end{tabular}
\caption{
    Magnitude of hoppings (in meV) between orbitals at positions denoted by the orange markers in 
    Fig.~\ref{fig:shellmap}, for $\theta = 1.05^\circ$. Cells with two entries correspond to the two distinct sets of equidistant $C_3$-related neighbors (filled and empty markers in Fig.~\ref{fig:shellmap}) 
    that can have different hopping magnitudes.
    \label{tab:hoppings1}
    }
\end{table}

\subsection{Symmetrized hoppings}

We provide the dominant hopping parameters of the symmetrized ten-band model $H_{\rm proj}$ at $\theta = 1.05^{\circ}$ in Tables~\ref{tab:hoppings0} and~\ref{tab:hoppings1},
with the full set of hopping parameters for angles down to $0.93^\circ$ available in~\cite{DATA}.
The neighbor shells and $C_3$ orbits that these tables refer to are shown in Fig.~\ref{fig:shellmap}.

\subsection{Symmetrization of the orbital basis}

In a given valley, each symmetrized orbital $\psi_b(\rv)$ is a linear combination of the ten \textsc{Wannier90} orbitals $w_n(\rv)$, 
\begin{equation}
    \psi_b(\rv) = \sum_{n=1}^{10} W_{nb} \, w_n(\rv), \qquad b = 1, \ldots, 10,
    \label{eq:symmorbitals}
\end{equation}
where $W = Q\Phi$ is assembled from the two basis changes: the rotation $Q$ into the eigenbasis of the fitted $D(C_3)$, which gives every orbital a $C_3$ eigenvalue $\omega^{m_b}$, and the transformation $\Phi$ of Eq.~\eqref{eq:pobasis}, which brings $D(C_{2y})$ and $D(C_2\mathcal{T})$ to the form of Table~\ref{tab:characters}.
The ten orbitals that result have the symmetry eigenvalues corresponding to the representations of the left-hand side of Eq.~\eqref{eq:repnew_appendix}.
The symmetrization of the Hamiltonian therefore reproduces the representation content of Ref.~\cite{po2019faithful} with no input regarding the shapes of the orbitals.

\clearpage 

\end{document}